\documentclass[conference]{IEEEtran}
\IEEEoverridecommandlockouts
\usepackage{cite}
\usepackage{amsmath,amssymb,amsfonts}
\usepackage{algorithmic}
\usepackage{graphicx}
\usepackage{subcaption}
\usepackage{textcomp}
\usepackage{xcolor}
\usepackage{multirow}
\usepackage{algorithm}
\usepackage{subcaption}
\usepackage{graphicx}
\usepackage{mathtools}
\usepackage{dsfont}
\DeclareMathOperator*{\argmax}{arg\,max}

\DeclarePairedDelimiter\floor{\lfloor}{\rfloor}
\usepackage[letterpaper, left=0.625in, right=0.625in, bottom=1.02in, top=0.70in]{geometry}
\begin{document}

\title{Backdoor Attacks and Defenses on Semantic-Symbol Reconstruction in Semantic Communications
% {\footnotesize \textsuperscript{*}}
}

\author{\IEEEauthorblockN{Yuan Zhou*, Rose Qingyang Hu*, Yi Qian$^\dag$}

\IEEEauthorblockA{
*Department of Electrical and Computer Engineering, Utah State University, Logan, UT, USA\\
$^\dag$Department of Electrical and Computer Engineering, University of Nebraska-Lincoln, NE, USA\\
Email: *\{yuan.zhou@ieee.org, rose.hu@usu.edu\}, $^\dag$yi.qian@unl.edu}}
\maketitle

\begin{abstract}
Semantic communication is of crucial importance for the next-generation wireless communication networks. The existing works have developed semantic communication frameworks based on deep learning. However, systems powered by deep learning are vulnerable to threats such as backdoor attacks and adversarial attacks. This paper delves into backdoor attacks targeting deep learning-enabled semantic communication systems. Since current works on backdoor attacks are not tailored for semantic communication scenarios, a new backdoor attack paradigm on semantic symbols (BASS) is introduced, based on which the corresponding defense measures are designed. Specifically, a training framework is proposed to prevent BASS. Additionally, reverse engineering-based and pruning-based defense strategies are designed to protect against backdoor attacks in semantic communication. Simulation results demonstrate the effectiveness of both the proposed attack paradigm and the defense strategies.
\end{abstract}


\section{Introduction}
Three communication levels established 70 years ago, namely, symbol transmission, semantic exchange, and the effects of semantic exchange, have outlined three pivotal issues in communications \cite{ex1}. The symbol transmission level mainly focuses on the accuracy of physical layer transmission of symbols or bits, which has been extensively explored in the past several decades. However, the communication frameworks based on Shannon paradigm have been gradually approaching to their theoretical limit, which results in the dilemma that wider and wider bandwidth is needed to support the exponentially increasing traffic volume. At the same time, we are encountering increasingly pressing challenges related to spectrum scarcity and the utilization of high-frequency bandwidth, particularly for outdoor coverage and mobility scenarios. Moving towards the sixth generation (6G) wireless networks, semantic communication defined in the second and third levels of communications has been proposed and envisioned as a promising technique to address the bandwidth and spectrum issues \cite{ref20}, \cite{ref21}. 

% Different from the traditional communication systems based on Shannon theorem that aim at accurately transmitting bit stream, semantic communication focusing on semantic exchange level aims to convey semantic information effectively by taking advantage of the background knowledge of the environment and the tasks. Moreover, semantic communication makes it possible to realize network intelligence by understanding the meaning of the transmitting message. {\color{blue}It is envisioned  that 6G will support massive immersive communications by integrating the digital and real worlds and introducing intelligence as part of the wireless communication systems, which highlights the key role of semantic communication \cite{ref21}.} %As a promising technique beyond Shannon paradigm, the semantic communication has received ever increasing attention both from academia and industrial.
% It is worth noting that the majority of existing semantic communication frameworks are based on deep learning due to its capability of extracting abstract features and learning context knowledge from raw data \cite{ref2, ref1}. However, deep learning empowered systems are vulnerable to attacks against learning systems, among which backdoor attack is one of the most commonly mentioned threats \cite{ex5}. 
It is worth noting that the majority of existing semantic communication frameworks are based on deep learning. Deep learning-based systems are vulnerable to attacks against learning systems, among which backdoor attack is one of the most commonly mentioned threats \cite{ex5}. The goal of traditional backdoor attacks is to intentionally deceive the target model to classify the poisoned data into adversary-specified class while preserving the original performance of the model with clean inputs. To achieve this goal, the adversary is able to poison training dataset \cite{ref8}. In existing works on backdoor attacks in wireless communication, the backdoor mainly targets the downstream classification model. In \cite{ref11}, a backdoor attack against wireless signal classifiers was developed, where the triggers are the signals with modified phase. Additionally, the investigation of backdoor attacks in the context of semantic communication was explored in \cite{ref6}. Nevertheless, existing backdoor attacks on semantic communication cannot be applied to semantic communication tasks with high-dimensional outputs, such as the transmission of images and speech, and semantic segmentation. Furthermore, this type of backdoor attacks can be detected and mitigated by existing defenses that have been fully developed for backdoor attacks against deep learning models. This paper focuses on the backdoor attacks in semantic communication with the capability to manipulate the semantics of reconstructed symbols in semantic communication. The backdoor attacks on semantic symbols (BASS) is proposed first, followed by an analysis on the defense methods against the traditional backdoor attacks. Building upon that, we leverage the distinct characteristics of BASS to propose three defense methods against the backdoor attacks. A training framework is first proposed to prevent data alteration. The second defense mechanism is based on reverse engineering to find the backdoor trigger. The last defense strategy focuses on mitigating the backdoor on models by pruning backdoor related neurons.

\begin{figure*}[htbp] % the asterisk makes the figure span both columns
    \centering
    
    % First figure inside a minipage
    \begin{minipage}{0.32\textwidth}
        \includegraphics[width=\linewidth]{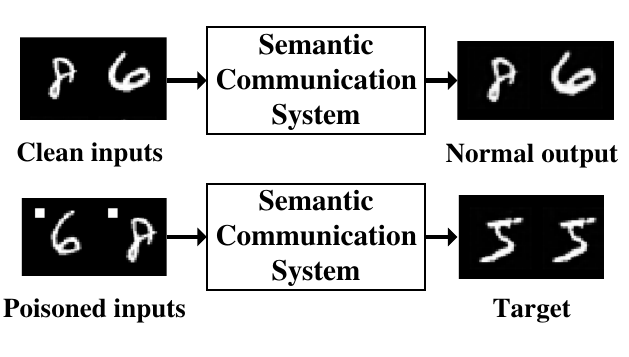}
        \caption{Comparison between the outputs of the backdoor model with poisoned and clean inputs.}
        \label{fig:backdoor_outpus}
    \end{minipage}
    \hfill % Spacing between the two figures
    % Second figure inside a minipage
    \begin{minipage}{0.66\textwidth}
        \includegraphics[width=\linewidth]{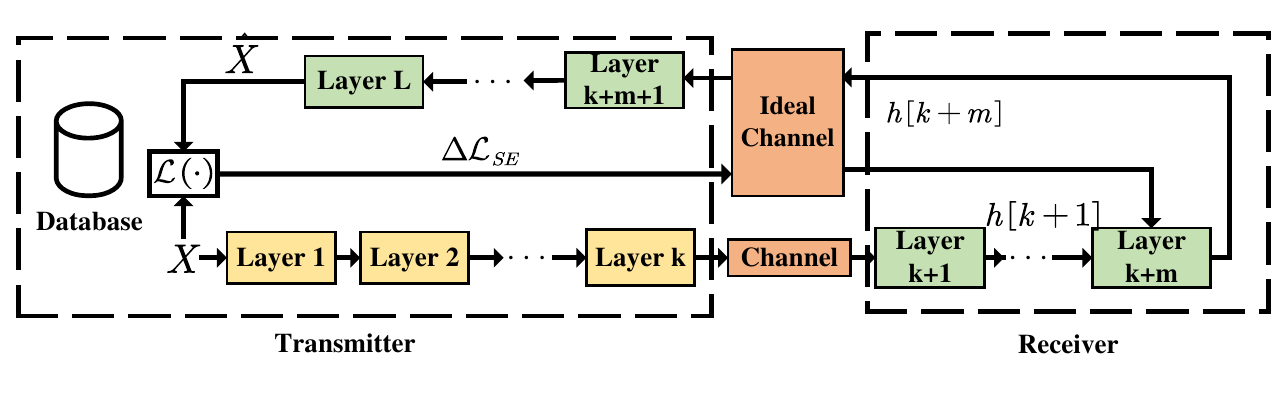}
        \caption{Training framework for preventing the BASS.}
        \label{fig:training_framework}
    \end{minipage}
\vspace{-0.8cm}
\end{figure*}
The rest of the paper is organized as follows. Section II presents the system model, alongside the threat model and the semantic communication framework. Section III defines a new backdoor attack paradigm targeting semantic communication systems. Defensive strategies against these attacks are detailed in Section IV. Simulation results are provided in Section V, and Section VI offers the conclusion.

% In Section III, a backdoor attack against semantic communication systems is proposed. Two adversarial attack methods are presented in Section IV. Defense schemes against the proposed attacks are elaborated  in Section V. Section VI gives the simulation results and Section VII concludes the paper.

\section{System model and preliminaries}
\subsection{System model}
A semantic communication system consisting of one transmitter and one receiver is considered. The training dataset for the symbols intended for transmission is located at the transmitter, while the dataset for the symbols to be recovered, along with their corresponding labels, is situated at the receiver.

In a typical neural network for semantic communication, there are several key components: the semantic encoder $\mathbf{S}_{\beta} (\cdot)$, the channel encoder $\mathbf{S}_{\alpha} (\cdot)$, the physical channel, the channel decoder $\mathbf{C}_{\delta}^{-1} (\cdot)$, and the semantic decoder $\mathbf{C}_{\mathcal{X}}^{-1} (\cdot)$. Here, $\beta$, $\alpha$, $\delta$, and $\mathcal{X}$ represent the parameters of the semantic encoder, the channel encoder, the channel decoder, and the semantic decoder, respectively.

\subsection{Threat model}

An attacker possesses the capability to manipulate both the training datasets at the receiver and the inputs at the transmitter. Specifically, the adversary can introduce triggers to modify the semantic symbols at the transmitter and also alter symbols in the receiver's dataset in accordance with the targets and labels specified by itself.

% \subsection{Semantic communication framework}
% The widely used convolutional neural network (CNN) based semantic communication model presented in \cite{ref5} is considered here. A typical semantic communication network consists of the semantic encoder $\math{S}_{\beta}(\cdot)$, the channel encoder $\math{C}_{\alpha}(\cdot)$, the physical channel, the channel decoder $\math{C}_{\delta}^{-1}(\cdot)$, and the semantic decoder $\math{C}_{\mathcal{X}}^{-1}(\cdot)$, where $\beta$, $\alpha$, $\delta$, and $\mathcal{X}$ are the parameters of the semantic encoder, the channel encoder, the channel decoder, and the semantic decoder, respectively. %The extracted features of the sentences are transmitted through wireless channel and the sentences are recovered from the received signals.
% % The $i$th sample of the training dataset is denoted as $\mathbf{x}_i, i=\{1,2,\cdots,D\}$. 

% The mean squared error (MSE) is used in the reconstruction loss, which is given by
% \begin{eqnarray}
% \mathcal{L}(\mathbf{x},\mathbf{\hat{x}})=\sum_{i=1}^{D} (\mathbf{x}_i-\mathbf{\hat{x}}_i)^2.
% \end{eqnarray}

% the loss function for the transmission task is cross-entropy, which is given by
% \begin{eqnarray}
% \begin{split}
% \mathcal{L}(\mathbf{y},\mathbf{\hat{y}})=&
% -\displaystyle\sum_{\maht{l=1}}^{\math{L}}\math{q(y_l)}\log(\math{p(y_l)})\\
% & +\displaystyle\sum_{\maht{l=1}}^{\math{L}}\maht{(1-q(y_l))}\log \math{(1-p(y_l))},
% \end{split}
% \end{eqnarray}
% where $q(\cdot)$ is the target probability of the words and $p(\cdot)$ is the predicted probability of the words.

\section{Backdoor Attacks Against Semantic Communication}
In this paper, we investigate a backdoor attack that can deceive the semantic communication system by manipulating the semantics of the reconstructed semantic symbols. The training datasets at the transmitter and the receiver are represented as $\mathcal{D}_T$ and $\mathcal{D}_R$ respectively. $\mathcal{D}_T$ comprises the semantic symbols $\{\mathbf{x}^T_i | i=1,2,\dots,N\}$ that are intended for transmission, whereas $\mathcal{D}_R=\{(\mathbf{x}^R_i, \mathbf{y}_i) | i=1,2,\dots,N\}$ includes the semantic symbols to be recovered and the labels of the semantic symbols. Specified triggers are introduced to a subset of $\mathcal{D}_T$, which is denoted as $\mathcal{D}_t$. Subsequently, the corresponding semantic symbols and the labels at the receiver are modified to achieve the intended semantics specified by the attack.
%  are represented as $\mathcal{Y}=\{\mathbf{y}_i | i=1,2,\dots,N\}$

During the training phase, the adversary introduces triggers into a specific proportion of input samples at the transmitter, turning them into poisoned samples. The semantic symbols and labels at the receiver are changed to match the target symbols and labels specified by the adversary. During the inference phase, the adversary injects triggers to poison the inputs, resulting in the produced reconstructed symbols being the target symbols specified by the adversary. Meanwhile, the model performs normally with clean inputs. Fig. \ref{fig:backdoor_outpus} shows how the attacked model works differently under poisoned inputs and under clean inputs.

\section{Defense Mechanism}
In contrast to conventional backdoor attacks, which primarily focus on manipulating classification outcomes, the proposed backdoor attacks aim to modify the reconstructed semantic symbols at the receiver. The semantic-symbol reconstruction in semantic communication involves high-dimensional semantic outputs, such as text, speech, and images. Furthermore, unlike conventional backdoor attack models where the activation patterns in poisoned and clean data can be separated, the defender in semantic communication is difficult to do so, even when the data distribution of the target semantic symbols differs from that of the clean data. Consequently, it is essential to investigate defense methods against BASS. Some widely accepted assumptions are made on the defender's capabilities, outlined as follows: Firstly, the training process is under the complete control of the defender. Secondly, the transmitter has unrestricted access to the entire training dataset $\mathcal{D}_T$, and likewise, the receiver also has unrestricted access to the complete training dataset $\mathcal{D}_T$. Thirdly, any portion of the training dataset can be targeted for attacking. Lastly, the defender has no access to any extra clean dataset.

% These defending methods rely on some widely accepted assumptions on the defender's capabilities, outlined as follows: First, the training process is under the complete control of the defender. Second, the transmitter has unrestricted access to the entire training dataset $\mathcal{D}_T$, and likewise, the receiver also has unrestricted access to the complete training dataset $\mathcal{D}_T$. Third, any portion of the training dataset can be targeted for attacking.

\subsection{Data location}
One crucial assumption for successful attack is that the transmitter is restricted to accessing only $\mathcal{D}_T$, while the receiver can only access $\mathcal{D}_R$. Based on the assumption, a semantic communication training framework is proposed to prevent BASS. Inspired by split learning, an U-shape forward-propagation and backward-propagation process is designed to push both $\mathcal{D}_T$ and $\mathcal{D}_R$ to the transmitter to prevent separate data poisoning. As shown in Fig. \ref{fig:training_framework}, the forward propagation initiates at the transmitter and ends at the same point. After passing through the encoder and wireless channel, the receiver continues the forward propagation until the $(k+m)$th layer to produce the activation signal $h[k+m]$, which is sent back to the transmitter through an "ideal channel". %The channel is an ideal communication medium, provided by the surrounding base stations just during the training phase. 
The transmitter ends the forward propagation and calculates the gradients, which are then sent to the receiver to update the parameters. This is done through the backward propagation. %{\color{red}As shown in Fig. 1.}

\subsection{Reverse engineering}
Among the defense methods proposed to counter traditional backdoor attacks,  reverse engineering is a commonly utilized approach to estimate the adversary's trigger pattern. The discrete outputs in conventional backdoor attacks enable the estimation of a trigger that can induce mis-classification with minimal input modification. However, pinpointing a specific target in a continuous space as a potential target is challenging.

The general form of triggering injection is assumed as follows,
\begin{equation}
  % \begin{split}
\mathbf{x}'_k = \mathbf{m}\cdot \mathbf{x}_k+(1-\mathbf{m})\cdot \Delta,
\end{equation}
Here, $\mathbf{m}$ represents a matrix that determines the extent to which the original image can be overwritten by the trigger, and $\Delta$ denotes the trigger pattern to be revealed, with the same dimension as the input image. These two estimation variables are optimized to discover the trigger, with values of $\mathbf{m}$ and $\Delta$ spanning from 0 to 1.

In BASS, the semantic feature distance between two poisoned samples with the same target should be smaller than that between two clean samples. Meanwhile, the backdoor can be activated with a small trigger. With $(\mathbf{m}, \Delta)$ defined as the variables of the trigger to be optimized, it can be estimated by minimizing the level of which the original images are overwritten by the trigger and the distance of the semantic features between two samples that are poisoned with the estimated labels. This process of estimating the trigger pattern can be formulated as the following optimization problem.
\begin{equation}
  \begin{split}
   \ \ \ \ \mathbf{P}_1 \ \min_{ \substack{ \{\mathbf{m}\}, \{\Delta\} }} &  \ \sum_{k\neq j}  \Vert \mathbf{En}(\mathbf{m} \cdot \mathbf{x}_k + (1-\mathbf{m})  \cdot 
 \Delta ) -\\ & \ \ \ \ \mathbf{En}(\mathbf{m} \cdot \mathbf{x}_{j} + (1-\mathbf{m}) \cdot \Delta ) \Vert^{2} + \lambda \Vert \mathbf{m} \Vert   \nonumber \\
    \text{s.t. } &
    \begin{aligned}[t]
    \ \ &\mathbf{0}<\mathbf{m} <\mathbf{1} \ \ \text{and} 
     &\mathbf{0}<\Delta<\mathbf{1}. \nonumber
    \end{aligned}
  \end{split}
\end{equation}
% The above optimization problem can solved by using penalty method. %In simulation, reverse engineering performs effectively on basic datasets such as MNIST, but it may converge to alternative adversarial patterns when applied to more intricate datasets. For different poisoned inputs, the poisoned model alters the semantics of the inputs to align with the same target's semantics. To estimate the trigger pattern, the first term in $\mathbf{P}_1$ that computed by pairs of poisoned data with the real trigger should be small. However, if the model's capacity is inadequate for modifying the semantics of the inputs by the encoder to get close semantic features, the first term of $\mathbf{P}_1$ can be great and thus cause failure.  %{\color{red}It is possible to force the semantic communication model to complete the processing of the poisoned data by adding an aid term to the loss, which will be explored in our future work.}

\begin{table*}[!t]
\vspace{-0.02cm}
\begin{center}
\caption{Average performance of the attack over SNRs}
\begin{tabular}{|c|cccc|cccc|}
\hline
\multirow{3}{*}{\begin{tabular}[c]{@{}c@{}}\textbf{Poison}\\ \textbf{Ratio}\end{tabular}} & \multicolumn{4}{c|}{\textbf{MNIST}}                                                                                        & \multicolumn{4}{c|}{\textbf{CIFAR10}}                                                                                     \\ \cline{2-9} 
                                                                        & \multicolumn{2}{c|}{\textbf{\textit{CR=1/8}}}                             & \multicolumn{2}{c|}{\textbf{\textit{CR=1/4}}}                             & \multicolumn{2}{c|}{\textbf{\textit{CR=1/8}}}                             & \multicolumn{2}{c|}{\textbf{\textit{CR=1/4}}}                             \\ \cline{2-9} 
                                                                        & \multicolumn{1}{l|}{\textbf{\textit{PSNRC}}} & \multicolumn{1}{l|}{\textbf{\textit{PSNRP}}} & \multicolumn{1}{l|}{\textbf{\textit{PSNRC}}} & \multicolumn{1}{l|}{\textbf{\textit{PSNRP}}} & \multicolumn{1}{l|}{\textbf{\textit{PSNRC}}} & \multicolumn{1}{l|}{\textbf{\textit{PSNRP}}} & \multicolumn{1}{l|}{\textbf{\textit{PSNRC}}} & \multicolumn{1}{l|}{\textbf{\textit{PSNRP}}} \\ \hline
0.01                                                                    & \multicolumn{1}{c|}{28.310}     & \multicolumn{1}{c|}{29.507}     & \multicolumn{1}{c|}{26.225}     & 31.028                         & \multicolumn{1}{c|}{25.948}     & \multicolumn{1}{c|}{13.818}     & \multicolumn{1}{c|}{27.042}     & 9.503                          \\ \hline
0.1                                                                     & \multicolumn{1}{c|}{27.543}     & \multicolumn{1}{c|}{26.605}    & \multicolumn{1}{c|}{26.048}    & 26.781                         & \multicolumn{1}{c|}{25.694}    & \multicolumn{1}{c|}{29.966}    & \multicolumn{1}{c|}{26.556}    & 30.665                         \\ \hline
0.2                                                                     & \multicolumn{1}{c|}{27.475}    & \multicolumn{1}{c|}{26.071}    & \multicolumn{1}{c|}{25.759}    & 25.211                    & \multicolumn{1}{c|}{24.260}    & \multicolumn{1}{c|}{30.975}    & \multicolumn{1}{c|}{25.206}    & 33.482                        \\ \hline
0.3                                                                     & \multicolumn{1}{c|}{26.874}    & \multicolumn{1}{c|}{26.166}    & \multicolumn{1}{c|}{25.274}    & 28.733                         & \multicolumn{1}{c|}{24.065}    & \multicolumn{1}{c|}{31.739}    & \multicolumn{1}{c|}{25.025}    & 33.907                         \\ \hline
0.4                                                                     & \multicolumn{1}{c|}{25.731}    & \multicolumn{1}{c|}{26.827}    & \multicolumn{1}{c|}{24.838}    & 28.733                         & \multicolumn{1}{c|}{23.251}    & \multicolumn{1}{c|}{33.133}    & \multicolumn{1}{c|}{24.632}    & 33.940                        \\ \hline
\end{tabular}
\end{center}
\vspace{-0.6cm}
\end{table*}
\subsection{Post-training pruning-based algorithm}
In semantic communication, semantic features are extracted by the encoder, while the decoder reconstructs the transmitted symbols. When the backdoor is activated, the backdoor model alters the semantics of the inputs to match a specified target. While the backdoor model retains and restores the semantics of the clean inputs, it discards the semantics of the poisoned inputs except for the trigger portion in different samples. These two contrary operations are executed by different neurons. Based on this fact, a pruning method is proposed to eliminate the backdoor by pruning the neurons in the encoder of the semantic communication network. The pruning operation is confined to the encoder for two primary reasons: First, it targets the neurons that activating the backdoor in the encoder; and second, it preserves the decoder's reconstruction capability, thereby minimizing the impact of pruning.

% In semantic communication, the encoder compresses the semantic symbols and the decoder recovers the transmitted symbols. Once the backdoor is activated, the semantics of the inputs are changed to that of the specified target. The semantics of the clean inputs are preserved and restored as much as possible while the semantics of the original inputs are discarded. These two contrary operations are executed by different neurons. 
Since convolutional networks are the predominant structure for image transmission in semantic communication, we focus on pruning the feature kernels of convolutional layers. This method can also be adapted to semantic communication frameworks based on other neural networks. This adaptation can be achieved by tailoring the pruning method to the specific requirements of the corresponding deep learning model.

Denote the parameters of the encoder with $L$ convolutional layers as $\mathcal{W}=\{(\mathbf{w}_1,b_1),(\mathbf{w}_2,b_2),\dots,(\mathbf{w}_L,b_L)\}$, where $\mathbf{w}_\ell\in\mathbb{R}^{f_{\ell}\times f_{\ell-1}\times H \times W}$ is the weight of the $\ell$th layer's convolutional kernels, $\mathbf{b}_\ell\in\mathbb{R}^{f_{\ell}}$ is the bias of the $\ell$th layer's convolutional kernels, and $f_\ell$ is the number of output channels in the $\ell$th layer, $f_0$ is the number of input channels. The size of the convolutional kernel is $H \times W$. %The feature map of the $l$th layer is represented as $\mathbf{z}_{l}^{k}\in\mathbb{R}^{H_{l},W_{l}}$, where $H_{l}$ and $W_{l}$ are its height and width, respectively. 

The objective of the pruning method is to eliminate the backdoor while simultaneously preserving the reconstruction accuracy of the semantic communication model, formulated as 
\begin{equation}
\label{p2}
  \begin{split}
   \mathbf{P}_2 \ \min_{ \substack{ \mathcal{W}}} & (\mathcal{C}(\mathcal{D}_{PC}|\mathcal{W})  - \mathcal{C}(\mathcal{D}_{P}|\mathcal{W}')) \\ & + \gamma |\mathcal{C}(\mathcal{D}_{C}|\mathcal{W}')-\mathcal{C}(\mathcal{D}_{C}|\mathcal{W})|. \nonumber
  \end{split}
\end{equation}
% \begin{equation}
%   \begin{split}
%    \ \ \ \ \mathbf{P}_1 \ \max_{ \substack{ \{\mathbf{m}\}, \{\Delta\} }} &  \ \sum_{k\neq j}  \Vert \mathbf{En}(\mathbf{m} \mathbf{x}_k + (1-\mathbf{m}) \Delta ) -\\ & \ \ \ \ \mathbf{En}(\mathbf{m} \mathbf{x}_{j} + (1-\mathbf{m}) \Delta ) \Vert^{2} + \lambda \Vert \mathbf{m} \Vert   \nonumber \\
%     \text{s.t. } &
%     \begin{aligned}[t]
%     \ \ &\mathbf{0}<\mathbf{m} <\mathbf{1},  \nonumber\\
%     \ \ &\mathbf{0}<\Delta<\mathbf{1}. \nonumber
%     \end{aligned}
%   \end{split}
% \end{equation}
Here $\mathcal{C}(\cdot)$ is the reconstruction accuracy. The parameters before pruning are denoted as $\mathcal{W}$ and those after pruning are represented as $\mathcal{W}'$. $\mathcal{D}_{P}$ and $\mathcal{D}_{C}$ are the collections of poisoned data and clean data, respectively. $\mathcal{D}_{PC}$ is the original benign data corresponding to $\mathcal{D}_{P}$. $\gamma$ is the parameter used to achieve the balance between the accuracy degradation and the backdoor cancellation. To get the optimal solution of $\mathbf{P}_2$, the defender needs to distinguish all the clean data and the poisoned data from the training dataset. Additionally, the transmitter and the receiver need to have access to the original unpoisoned datasets corresponding to the poisoned data. However, these conditions are not feasible for the defender. Therefore, an algorithm is proposed to search for an approximate solution.

% The pruning-based algorithm eliminates the backdoor with several steps described as follows: (1) The network is first pruned to get the changes on semantic features for each sample. (2) Considering that the semantics of clean data remains stable when pruned, while the semantics of poisoned data undergoes significant changes during the pruning process, it is possible to distinguish the clean and the poisoned data within the sampled dataset. (3) Since the significant performance fluctuations of the poisoned samples during the pruning process, the optimal pruning ratio is determined by assessing the variations in semantic features of the poisoned data. 
% There are two problems that need to be solved. First, the clean data and poisoned data need to be recognized from the sampled subset of the training data. Second, the reconstruction accuracy of the poisoned data cannot be determined unless the receiver obtains the recovered semantic symbols. 

% The details of the proposed algorithm are presented in \textbf{Algorithm 1}. 
After training, the network is first pruned with different pruning ratios. Subsequently, the activations of the data are logged to identify the optimal pruning ratio. The first layer remain unpruned to prevent substantial performance degradation. The number of the pruned kernels is calculated, with one kernel being pruned at each iteration. Identifying an optimal subset of parameters while minimizing the deviation from the original cost value constitutes a challenging combinatorial problem. The number of pruned kernels of the $\ell$th layer is expressed as $\floor*{C_{out}^{\ell}r}+\mathds{1}(\ell)$, where $r$ is the pruning ratio, and $\floor*{\cdot}$ is the floor function. $C_{out}^{\ell}$ is the number of output feature maps of the $\ell$ layer. $\mathds{1}(\ell)$ is an indicator function defined as
\begin{equation}
\mathds{1}(\ell)=
\begin{cases}
1 \hspace{5mm} & \floor*{\sum_{\substack{i=\ell_0}}^{\substack{\ell_l}}{r C_{out}^{i}}} + \ell-\ell_0 < r \sum_{i=\ell_0}^{\ell_l}C_{out}^{i},\\ 0\hspace{5mm} & otherwise,
\end{cases}
\end{equation}

where $\ell_0\in\{1,2,...,L\}$ and $\ell_l\in\{2,...,L\}$ are the smallest and the largest indices of the pruned layers, respectively. In this paper, $\ell_0=2$, $\ell_l=4$. 

Identifying parameters associated with the backdoor involves analyzing the absolute sum of the elements of each feature map. To be specific, the median values of the absolute sums of feature maps are compared at each layer. Parameters associated with feature maps having lower median values are given priority in the pruning process. There are three possible types of useful parameters that can be pruned with this pruning method, namely, parameters related to the backdoor but unrelated to the model's normal functioning, parameters unrelated to the backdoor yet essential for the model's normal operation, and parameters tied to both the backdoor and normal model functions. The intuition for selecting the lowest sample median values is that the parameters associated with backdoor operations tend to produce high activation values with poisoned inputs, while these parameters yield lower activation values with clean inputs. As long as the proportion of the poisoned samples is smaller than $50\%$, the median values are dominated by clean data. The assumption of the poisoned samples is smaller than $50\%$ is reasonable. From the simulation results, the performance of the model can drop by more than $5\%$ when the poison ratio exceeds $40\%$. In addition, owing to the high-dimensional nature of semantic symbols, the target semantic symbols can be identified by matching samples in the training dataset at the receiver. If the similarity between two samples exceeds a predefined threshold, the pair of samples can be classified as poisoned. As such, BASS need to maintain a small poison ratio to guarantee its stealthiness. Consequently, the first case is most likely to occur when a small fraction of the training data are poisoned. If the second and third cases arise, the low activation values indicates that these parameters are relatively unimportant and will not substantially impact the model's performance.

By pruning the feature maps, the change of semantic features on $L^1$ norm $c_s^{q} = \frac{\Vert \mathbf{v}^{q}_s-\mathbf{v}^{q}_0\Vert}{\Vert\mathbf{v}^{q}_0\Vert}$ with different pruning ratios are logged, where the sampled pruning ratios are indexed by $s\in\mathbf{S}=\{0,1,2,\dots,S\}$. The model is unpruned when $s=0$. Here, $\mathbf{v}^{q}_{s}$ denotes the semantic feature of the $q$th sample in $\mathcal{D}'_T=\{\mathbf{x}_{1}^{t}, \mathbf{x}_{2}^{t},\dots, \mathbf{x}_{Q}^{t}\}$, where the model is with the pruning ratio indexed by $s$. And $\mathcal{D}'_T$ is a subset of $\mathcal{D}_T$, which is the subset of the training dataset at the transmitter side. %The number of poisoned samples in $\mathcal{D}'_T$ is denoted as $P$. 
In order to track the variation of the backdoor-related neurons as the pruning ratio increases, the poisoned samples should be identified. However, the poison ratio can be extremely small, which can cause data imbalance. To address the imbalance between poisoned and clean samples, a K-means model is trained with subsampled data from the training dataset. The samples in $\mathcal{D}'_t$ are first sorted by $c_s^{q}$. Then, the samples with $n\%$ greatest $c^{q}_{s}$ are classified as poisoned data and the samples with $n\%$ smallest $\mathbf{c}^p$ are clean data at $s=\argmax_{s} \frac{1}{Q} \sum_{q=1}^{Q} (c^{q}_{s}-\bar{c}_{s})^2$, where $\bar{c}_{s}=\frac{1}{Q}\sum_{q} {c}^{q}_{s}$. These $2n\%$ samples are used as training data of K-means. The sampled data are classified with the fitted K-means model.

In the simulation section, the results show that backdoor can be eliminated when the pruning ratio exceeds a certain point. Then, the performance with both clean and poisoned data gradually degrades. This process implies that the change on semantic features is first violent. Then, it becomes stable once the backdoor has been eliminated. Since the performance of clean data drops with the pruning ratio, the pruning ratio should be kept small whenever possible. Another observation is that the first term in $\mathcal{P}_2$ has no obvious decrease before the "knee", which along with the fact that the second term keep increasing with pruning ratio, the approximate optimal pruning point has to be appear at "knee" or $s=0$. Since the defender need to eliminate the backdoor while remaining the performance of the model, the optimal point has to be appear at the "knee". Thus the optimal pruning point can be found at the "knee" by employing the algorithm proposed in \cite{knee}. 

The curve $\mathbf{c}^p=[c_0^p, c_1^p, c_2^p, \dots, c_S^p]$ is utilized to find the best pruning ratio point, where $c_s^p=\frac{1}{J} \sum_{q} \mathds{1}'(q) c_s^{q}$, where $J$ is the number of poisoned samples in $\mathcal{D}'_T$. The indicator function $\mathds{1}'(q)=1$ if the $q$th sample is classified as poisoned sample. Conversely, $\mathds{1}'(q)=0$ if the $q$th sample is identified as clean sample. A sliding window of size $w=3$ is used to enhance the robustness. $\frac{w-1}{2}$ $0$s are appended to the head and tail of $\mathbf{c}^p$ to produce $\widetilde{\mathbf{c}}^p = [c_{-1}^{p}, c_{0}^{p}, \dots, c_{S+1}^{p}]$. The average with the sliding window is first calculated. $\mathbf{\Bar{c}}=[\Bar{c}_0,\Bar{c}_1,\dots,\Bar{c}_S]$, where $\Bar{c}_{s+1} = \frac{1}{w}\sum_{i=0}^{w} c_{s+i}, s=\{-1,0,\dots,S-1\}$. Then, the difference $d_c=norm(\mathbf{\Bar{c}})-norm(pr\_list)$ is calculated, where $norm(\cdot)$ represents the max-min normalization and $pr\_list$ is the vector consisting of pruning ratios. Let $d_c(pr)$ denotes the element of $d_c$ corresponding to the pruning ratio $pr$. The optimal pruning ratio $pr^{*}$ is determined by $\operatorname{argmax}_{pr} d_c (pr)$. To deal with the randomness in data sampling, executing the algorithm multiple iterations or increasing the number of samples serves to improve the method's robustness.

\section{Simulation}
In this section, the effectiveness of the proposed attack and defending methods are evaluated under different parameter settings. The backdoor models with spectrum ratio of $1/4$ and $1/8$ trained by MNIST and CIFAR10 are evaluated. We then maintain the poison ratio constant while varying the power of Gaussian noise and perturbations to evaluate performance across different Signal-to-Noise Ratios (SNRs) ranging from $1$ to $13$ dB. To evaluate the performance with different poison ratios, poison ratios of $0.01$, $0.05$, $0.1$, $0.2$, $0.3$, and $0.4$ are tested. The sub-sampling ratio for k-means is $n = 0.02$. All the models are trained for $120$ epochs with a learning rate of $0.0008$. And the number of the data points sampled for defense is $2000$. Additive White Gaussian Noise (AWGN) channel is considered in the following simulation runs. MNIST and CIFAR10 are used as the training datasets. To comprehensively evaluate the effectiveness of both attack and defense methods, we assess their performance in two different scenarios: 1) The target distribution is the same as the training dataset distribution; 2) The target distribution is different from the training dataset distribution. To be specific, the target is chosen from the MNIST training dataset in the first case. Consequently, the target distribution aligns with the distribution of the training dataset when the training dataset is MNIST. When the training dataset is CIFAR10, the target originates from an entirely different distribution. The peak signal-to-noise ratio (PSNR) is employed to measure the reconstruction accuracy. $\text{PSNR} = 10\log_{10}(\frac{R^2}{MSE})$, where $R$ is the maximum fluctuation in the images, and MSE is mean-squared error calculated by the reconstructed data of the decoder model and the target data.
\begin{figure}[htb]
\vspace{-0.8cm}
	\centering
 \begin{minipage}{0.48\textwidth}
 
\includegraphics[width=3.4in]{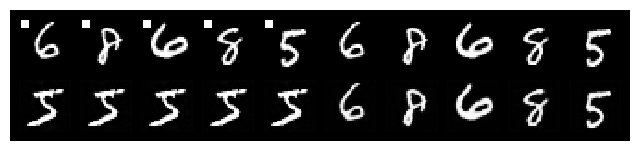}
	% \label{MNIST_ATTACK}
	
\end{minipage}
 \begin{minipage}{0.48\textwidth}
 
\includegraphics[width=3.4in]{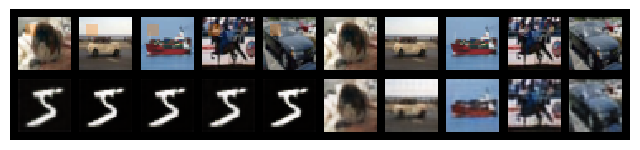}
	% \label{CIFAR_ATTACK}
	\caption{Inputs and reconstructed images of poisoned model with poisoned inputs and clean inputs with training dataset of MNIST and CIFAR10.}
\end{minipage}
\vspace{-0.5cm}
\end{figure}

Table 1 shows the average PSNR of clean (PSNRC) data and that of the poisoned (PSNRP) data when the poisoned model changes with poisoned ratios and compression ratios. As the poison ratio increases, the reconstruction accuracy for un-poisoned samples decreases. The dropping is attributed to the reduction in training data for normal function with the increase of poison ratio. Notably, the attack performance of MNIST may not consistently rise with the increase of poison ratio. 
% There are thresholds or tipping points in the behavior of the model. Initially, as the poison ratio increases, the model's performance may degrade. However, beyond a certain point, it adapts to the backdoor trigger more effectively, leading to an increase in PSNRP.

Fig. 3 shows the transmitted images and the reconstructed images of the poisoned models trained by MNIST and CIFAR10, respectively. The first row and the third row are the inputs of the backdoor semantic communication model. The second row and the fourth row are the corresponding reconstructed images of at the receiver. For the model trained by MNIST, a white square at the upper-left is added to the image while a colored square is used as trigger for CIFAR10. The outputs of the poisoned model are "5"s specified by the adversary and the model performs normal with clean data. 

\setlength{\abovecaptionskip}{0.cm}
\begin{figure}[htb]
\vspace{-0.2cm}
	\centering
	\includegraphics[width=2.8in]{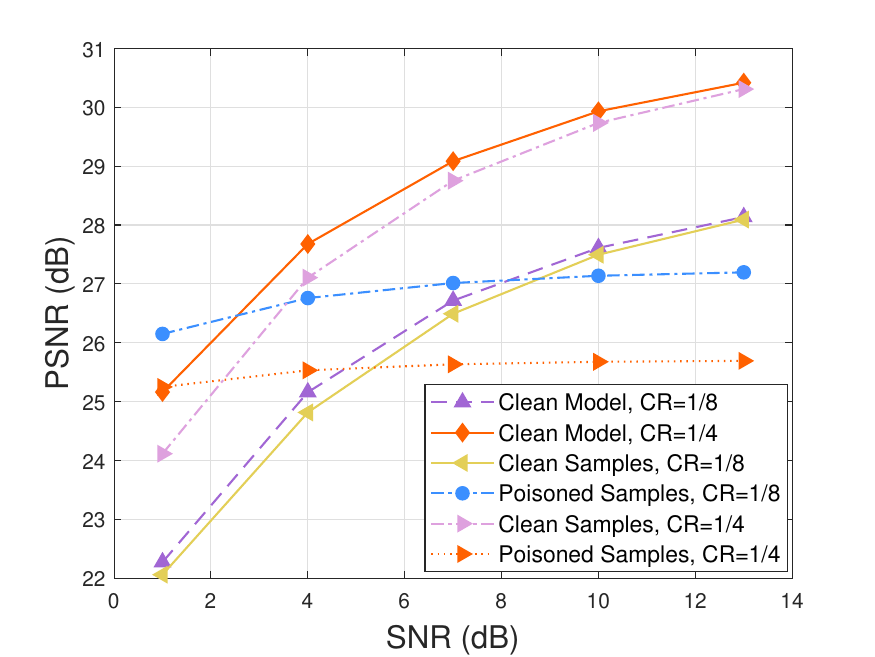}
	\label{Backdoor Model Loss1}
	\caption{PSNR comparison for models trained with MNIST at a $5\%$ poisoned ratio across different SNR levels and compression ratios.}
\vspace{-0.5cm}
\end{figure}

\setlength{\abovecaptionskip}{0.cm}
\begin{figure}[htb]
\vspace{-0.07cm}
	\centering
	\includegraphics[width=2.8in]{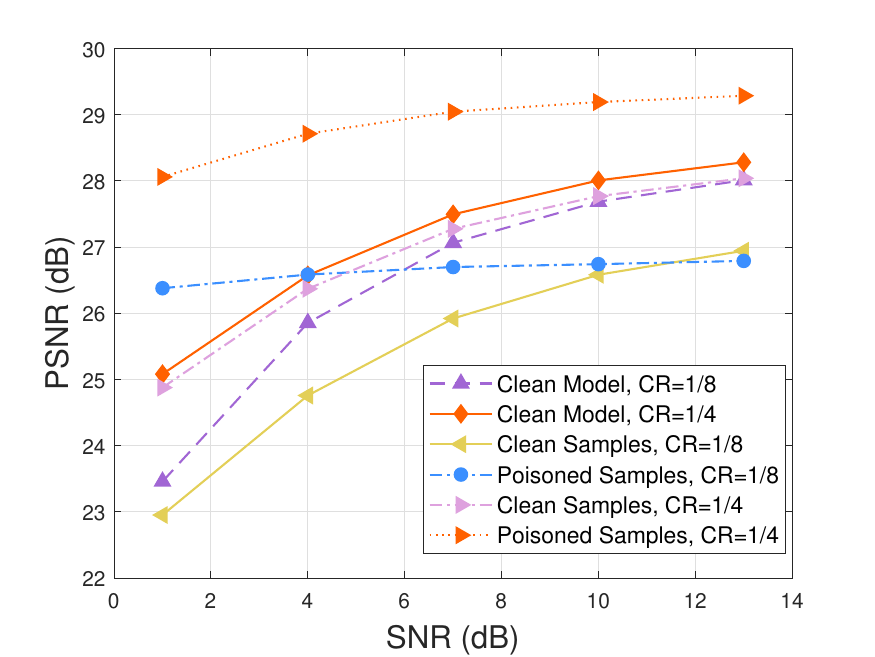}
	\label{Backdoor Model Loss2}
	\caption{PSNR comparison for models trained with CIFAR10 at a $5\%$ poisoned ratio across different SNR levels.}
\vspace{-0.5cm}
\end{figure}

Fig. 4 and Fig. 5 depict the comparison for the performance of backdoor and clean models trained with MNIST and CIFAR10 at a $5\%$ poisoned ratio across different SNR levels and compression ratios. Both the reconstructed quality for un-poisoned samples and the performance of the attack improve as SNR and compression ratio increase, which implies that BASS is more effective when the compression ratio and SNR are high. Another observation is that the reconstruction quality of the backdoor model with clean data is close to that of the clean model, highlighting that the attack remains highly effective in target samples without affecting the performance on un-poisoned samples. Notably, the reconstruction quality of clean data drops more than other cases in Fig. 5 where the models trained with CIFAR10 with compression ratio of $1/8$. It shows that the performance of the model on clean data can have more significant drops when the compression ratio is small. 
%In addition, the recovered performance of poisoned data in figure (b) is close to that of clean data, while the recovered performance of poisoned data in figure (a) achieves only about $70\%$ of clean data, which shows that the performance of poisoned data is difficult to recover when the target and the training data has the same distribution.
\begin{figure}
% \vspace{-0.25cm}
\begin{minipage}[t]{0.49\linewidth}
\begin{subfigure}{\textwidth}
    \centering
    \includegraphics[width=0.6in]{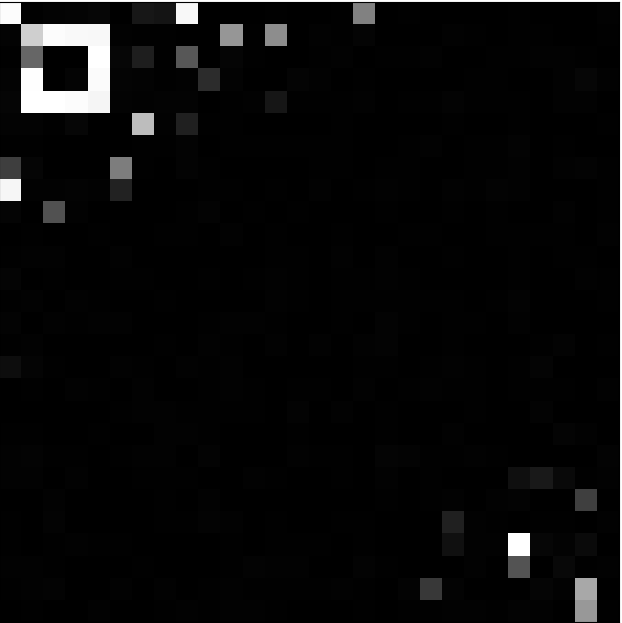}
    \caption{}
\end{subfigure}%
\end{minipage}
\begin{minipage}[t]{0.49\linewidth}
\begin{subfigure}{\textwidth}
    \centering
    \includegraphics[width=0.6in]{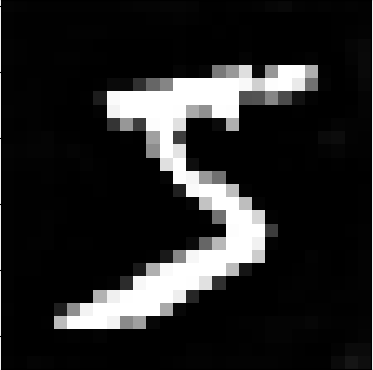}
    \caption{}
\end{subfigure}
\end{minipage}
\caption{(a) Trigger pattern estimated by the proposed method for attacks against MNIST. (b) The reconstructed output when the transmitted symbol is poisoned by the estimated trigger.}
\vspace{-0.7cm}
\end{figure}

Fig. 6 (a) shows the trigger pattern estimated by the proposed method for attacks against MNIST and the reconstructed output when the input is poisoned by the estimated trigger. Fig. 6 (b) shows that the backdoor is activated by the estimated trigger.
\begin{figure}
\begin{minipage}[t]{0.49\linewidth}
\begin{subfigure}{\textwidth}
    \centering
    \includegraphics[width=\linewidth]{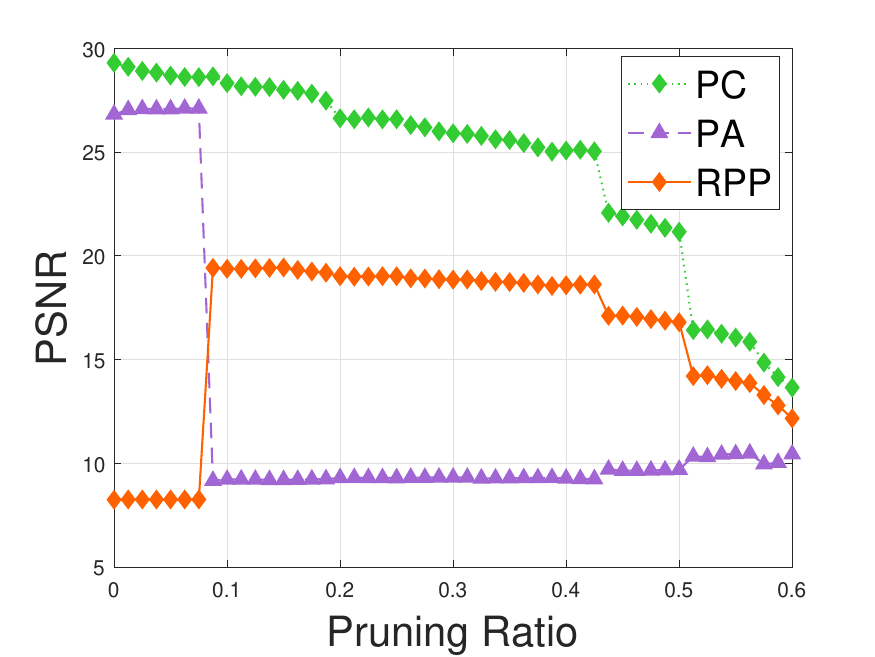}
    \caption{}
\end{subfigure}%
\end{minipage}
\begin{minipage}[t]{0.49\linewidth}
\begin{subfigure}{\textwidth}
    \centering
    \includegraphics[width=\linewidth]{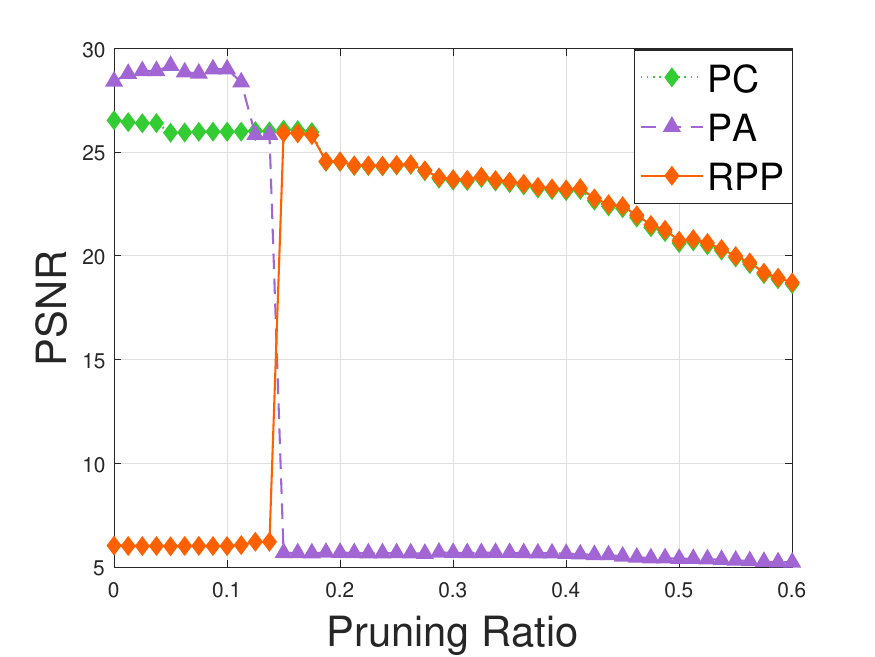}
    \caption{}
\end{subfigure}
\end{minipage}
\caption{Reconstruction accuracy of clean data and poison data versus pruning ratio. (a) PSNR versus pruning ratio with training dataset of MNIST. (b) PSNR versus pruning ratio with training dataset of CIFAR10.}
\vspace{-0.0cm}
\end{figure}

% The comparison of the mean activation values across the first four layers with different training dataset and target data are illustrated in Fig. 7 to Fig. 10. The results indicate a distinct correlation where activations significantly higher with the inputs of poison data correspond to feature maps that exhibit low activation values when processing clean samples. In other words, the orthogonality between the activations of clean and poison data makes it possible to eliminate the backdoor by pruning the parameters with low clean activation values. Furthermore, the low pruned activation values with clean data highlight the relatively minor significance of the parameters for normal operation. Another observation is that, with the exception of critical feature maps for poison data, activation values for poisoned data tend to decrease with increasing layer depth, in contrast to those observed for clean data. These observations corroborating our initial analysis on the activation patterns characteristic of BASS and provide foundational insights for understanding and defending the attack.

Fig. 7 illustrates the relationship between the pruning ratio and the reconstruction accuracy for both clean and poisoned data. It displays the performance of clean data (PC), the performance of the attack (PA), and the recovery performance of poisoned data (RPP). It can be observed that the reconstruction accuracy of the clean data remains stable, with a slight decrease when the pruning ratio is small. Conversely, the attack performance, which is measured by the PSNR of the reconstructed semantic symbols and the adversary-specified target, declines significantly when the pruning ratio increases. At certain pruning levels, the accuracy of poisoned data experiences a sharp rise. This trend suggests that by strategically pruning can mitigating backdoors, ensuring the model's behavior on poisoned data can be restored with minimal impact on the accuracy of clean data.

\setlength{\abovecaptionskip}{0.cm}

\begin{figure}[htb]
\vspace{-0.5cm}
	\centering
	\includegraphics[width=2.8in]{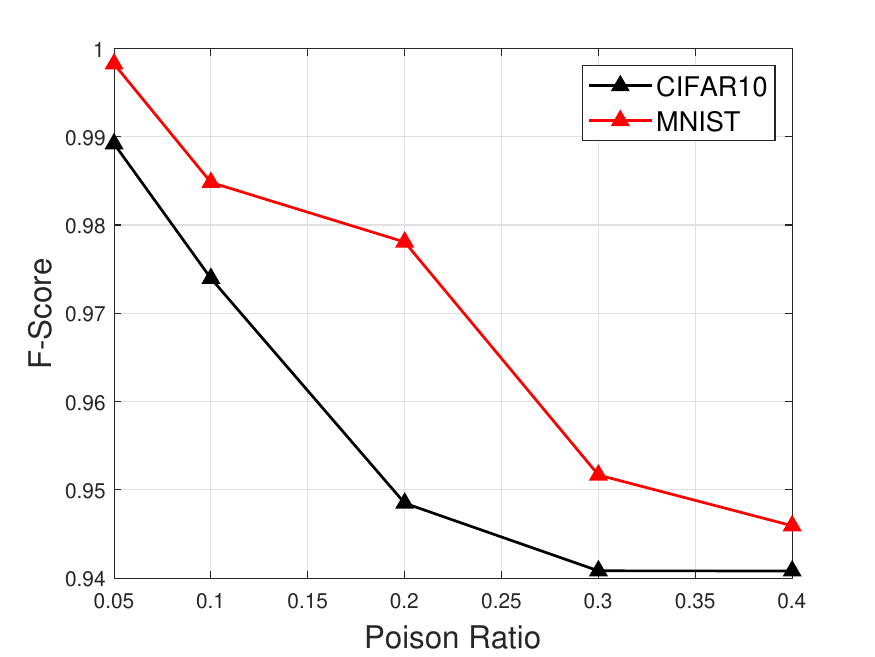}
	\label{Backdoor Model Loss}
	\caption{F1-score comparison for models trained with MNIST and CIFAR10 across different poison ratios.}
\vspace{-0.45cm}
\end{figure}

In Fig. 8, the achieved f1-score is shown to evaluate the performance of the proposed poisoned data identification method across different poison ratios. It shows a high classification accuracy for the backdoor model trained with CIFAR10 and MNIST. The achieved classification accuracy decreases with the increase of the poison ratio. 
When the poison ratio increases, more parameters that are unrelated to the backdoor or parameters tied to both the backdoor and normal model functions are pruned by the median value based-pruning, which produces more mis-classified samples in the fitting data for k-means. Consequently, classification performance degrades when poison ratio goes up.

\setlength{\abovecaptionskip}{0.cm}

\begin{figure}[htb]
\vspace{-0.42cm}
	\centering
	\includegraphics[width=2.8in]{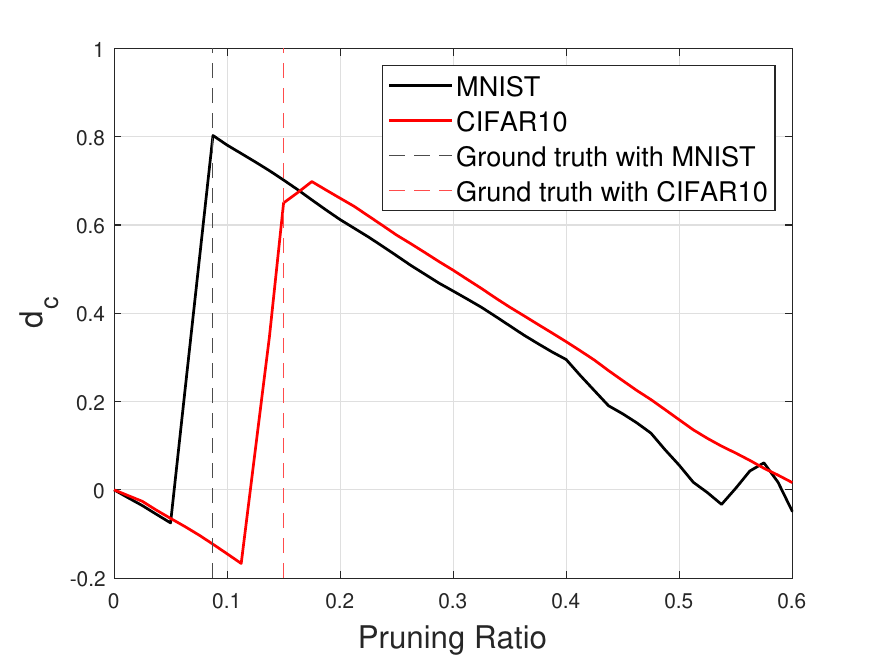}
	\label{dc}
	\caption{$d_c$ versus pruning ratio}
\vspace{-0.45cm}
\end{figure}

The difference $d_c$ is utilized to determine the optimal pruning ratio. The "ground truth" of the best pruning ratio point should be equal to or less than the pruning ratio determined by $d_c$ for successfully eliminating the backdoor. Fig. 9 demonstrates $d_c$ versus the pruning ratio when using the same experimental setup as in Fig. 7. In addition, the dashed line indicates the actual optimal pruning point. In both cases, the proposed method successfully eliminates the backdoors without noticeable performance degradation. The performance of clean data decreases by $2.224\%$ and $2.148\%$, respectively. Meanwhile, the reconstruction performance of poisoned data recovers to $67.791\%$ and $99.979\%$, respectively.

\section{Conclusions}
% Vulnerabilities of deep learning enabled semantic communication pose a significant safety risk. In this paper, we introduce a new paradigm of backdoor attacks on reconstructed symbols in semantic communication, which cannot be detected or mitigate by existing defending methods. As a countermeasure, a training framework is proposed to prevent the attack, a reverse engineering method is explored to estimate the trigger, and a pruning-based algorithm is proposed to eliminate the backdoor without reverse-training. Simulation results demonstrate the effectiveness of the proposed attacks and defense schemes.

This paper presents a novel paradigm of backdoor attacks targeting reconstructed symbols in semantic communication. The attacks cannot be detected or mitigated by the current defense strategies. To counter this threat, we propose a training framework designed to prevent such attacks. Additionally, we explore a reverse engineering approach for trigger estimation and design a pruning-based algorithm to eliminate the backdoor without re-training. Our simulation results validate the effectiveness of the proposed attack methods and defense strategies.

\section{Acknowledgment}
This work was partially supported by National Science Foundation under grants CNS-2008145, CNS-2007995, CNS-2319486, CNS-2319487.

\vspace{12pt}
\color{red}

\end{document}